\documentclass[prl,twocolumn,twoside,preprintnumbers,superscriptaddress,nofootinbib,showpacs]{revtex4-1}
\usepackage[colorlinks=true,citecolor=blue,filecolor=blue,linkcolor=blue,urlcolor=blue,pdftex]{hyperref}

\usepackage{xspace}
\usepackage[usenames,dvipsnames]{color}
\usepackage{booktabs,graphicx,mathrsfs,verbatim,amsmath,units,soul}
\usepackage{ulem}
\usepackage{xspace} 
\usepackage{upgreek}
\usepackage{textcomp}
\usepackage{gensymb}
\usepackage{tabularx}
\usepackage{xcolor}
\newcommand{\Sp}{\langle \mathbf{S}_\mathrm{p} \rangle}
\newcommand{\Sn}{\langle \mathbf{S}_\mathrm{n} \rangle}
\newcommand{\sigmaN}{\sigma^{\mathrm{SD}}_{\chi N}}
\newcommand{\sigman}{\sigma^{\mathrm{SD}}_{\chi n}}
\newcommand{\sigmap}{\sigma^{\mathrm{SD}}_{\chi p}}

\newcommand{\gevcsq}{GeV/c${}^2$}



\newcommand{\bologna}{\affiliation{Department of Physics and Astronomy, University of Bologna and INFN-Bologna, 40126 Bologna, Italy}}

\newcommand{\chicago}{\affiliation{Department of Physics \& Kavli Institute for Cosmological Physics, University of Chicago, Chicago, Illinois 60637, USA}}

\newcommand{\coimbra}{\affiliation{LIBPhys, Department of Physics, University of Coimbra, 3004-516 Coimbra, Portugal}}

\newcommand{\columbia}{\affiliation{Physics Department, Columbia University, New York, New York 10027, USA}}

\newcommand{\lngs}{\affiliation{INFN-Laboratori Nazionali del Gran Sasso and Gran Sasso Science Institute, 67100 L'Aquila, Italy}}

\newcommand{\mainz}{\affiliation{Institut f\"ur Physik \& Exzellenzcluster PRISMA, Johannes Gutenberg-Universit\"at Mainz, 55099 Mainz, Germany}}

\newcommand{\heidelberg}{\affiliation{Max-Planck-Institut f\"ur Kernphysik, 69117 Heidelberg, Germany}}

\newcommand{\munster}{\affiliation{Institut f\"ur Kernphysik, Westf\"alische Wilhelms-Universit\"at M\"unster, 48149 M\"unster, Germany}}

\newcommand{\nikhef}{\affiliation{Nikhef and the University of Amsterdam, Science Park, 1098XG Amsterdam, Netherlands}}

\newcommand{\nyuad}{\affiliation{New York University Abu Dhabi, Abu Dhabi, United Arab Emirates}}

\newcommand{\purdue}{\affiliation{Department of Physics and Astronomy, Purdue University, West Lafayette, Indiana 47907, USA}}

\newcommand{\rpi}{\affiliation{Department of Physics, Applied Physics and Astronomy, Rensselaer Polytechnic Institute, Troy, New York 12180, USA}}

\newcommand{\rice}{\affiliation{Department of Physics and Astronomy, Rice University, Houston, Texas 77005, USA}}

\newcommand{\stockholm}{\affiliation{Oskar Klein Centre, Department of Physics, Stockholm University, AlbaNova, Stockholm SE-10691, Sweden}}

\newcommand{\subatech}{\affiliation{SUBATECH, IMT Atlantique, CNRS/IN2P3, Universit\'e de Nantes, Nantes 44307, France}}

\newcommand{\torino}{\affiliation{INFN-Torino and Osservatorio Astrofisico di Torino, 10125 Torino, Italy}}

\newcommand{\ucla}{\affiliation{Physics \& Astronomy Department, University of California, Los Angeles, California 90095, USA}}

\newcommand{\ucsd}{\affiliation{Department of Physics, University of California, San Diego, California 92093, USA}}

\newcommand{\wis}{\affiliation{Department of Particle Physics and Astrophysics, Weizmann Institute of Science, Rehovot 7610001, Israel}}

\newcommand{\zurich}{\affiliation{Physik-Institut, University of Zurich, 8057  Zurich, Switzerland}}

\newcommand{\paris}{\affiliation{LPNHE, Universit\'{e} Pierre et Marie Curie, Universit\'{e} Paris Diderot, CNRS/IN2P3, Paris 75252, France}}

\newcommand{\freiburg}{\affiliation{Physikalisches Institut, Universit\"at Freiburg, 79104 Freiburg, Germany}}

\newcommand{\lal}{\affiliation{LAL, Universit\'e Paris-Sud, CNRS/IN2P3, Universit\'e Paris-Saclay, F-91405 Orsay, France}}

\newcommand{\naples}{\affiliation{Department of Physics ``Ettore Pancini'', University of Napoli and INFN-Napoli, 80126 Napoli, Italy}} 

\newcommand{\nagoya}{\affiliation{Kobayashi-Maskawa Institute for the Origin of Particles and the Universe, Nagoya University, Furo-cho, Chikusa-ku, Nagoya, Aichi 464-8602, Japan}}

\begin{document}

\title{Constraining the Spin-Dependent WIMP-Nucleon Cross Sections with XENON1T}

\author{E.~Aprile}\columbia
\author{J.~Aalbers}\stockholm\nikhef
\author{F.~Agostini}\bologna
\author{M.~Alfonsi}\mainz
\author{L.~Althueser}\munster
\author{F.~D.~Amaro}\coimbra
\author{M.~Anthony}\columbia
\author{V.~C.~Antochi}\stockholm
\author{F.~Arneodo}\nyuad
\author{L.~Baudis}\zurich
% ...
\collaboration{XENON Collaboration}
\email[]{xenon@lngs.infn.it}
\noaffiliation

\date{\today} 

\begin{abstract}
We report the first experimental results on spin-dependent elastic weakly interacting massive particle (WIMP) nucleon scattering from the XENON1T dark matter search experiment. The analysis uses the full ton year exposure of XENON1T to constrain the spin-dependent proton-only and neutron-only cases. 
No significant signal excess is observed, and a profile likelihood ratio analysis is used to set exclusion limits on the WIMP-nucleon interactions. This includes the most stringent constraint to date on the WIMP-neutron cross section, with a minimum of $6.3\times10^{-42}$~cm$^2$ at 30~\gevcsq~and 90\% confidence level. The results are compared with those from collider searches and used to exclude new parameter space in an isoscalar theory with an axial-vector mediator. 
\end{abstract}


\maketitle

{\it Introduction.}---There is a preponderance of astrophysical evidence that a nonluminous, massive component known as dark matter (DM) comprises about 26.5\% of the total energy density of the Universe~\cite{wimp_hooper, planck}. Still, the particle nature of this component remains unknown. 
One attractive DM candidate is the weakly interacting massive particle (WIMP), which arises naturally in several extensions of the standard model of particle physics~\cite{wimp_review, jungman_review}.
This has motivated many experimental searches for both the decay and self-annihilation products of WIMPs (indirect detection)~\cite{indirect}, for WIMP production at particle accelerators~\cite{lhc_review}, and for WIMPs scattering off atomic nuclei on Earth (direct detection)~\cite{laura_review}.
One leading direct detection technique uses liquid xenon (LXe) time projection chambers (TPCs), placed underground to reduce backgrounds induced by cosmic rays~\cite{xenon1t_sr1, lux_dm, panda_dm, xmass_dm}.
XENON1T, the largest and most sensitive of these experiments to date, is a dual-phase (liquid and gas) xenon TPC located at a depth of 3600~m water equivalent at the INFN Laboratori Nazionali del Gran Sasso in L'Aquila, Italy~\cite{xe_instrument}. 

XENON1T contains 3.2~t of ultrapure LXe, with 2~t in the active detector volume, and is outfitted with two arrays of 127 and 121 Hamamatsu R11410-21 3" photomultiplier tubes (PMTs)~\cite{xe_pmts, pmt2} facing the LXe from above and below, respectively.
Energy depositions in the active volume produce scintillation photons as well as ionization electrons. 
The scintillation light is promptly detected by the PMTs (S1), while the electrons are drifted upward through the LXe by a uniform electric drift field. At the liquid-gas interface, the electrons are extracted by an electric field towards the anode, producing proportional scintillation light (S2) via electroluminescence, which is also detected by the PMTs.
The time difference between the S1 and the S2 is proportional to the depth of the original interaction. Combined with the PMT hit pattern of the S2, this allows for 3D position reconstruction of the event.
The reconstructed position helps eliminate background events due to radioactivity from materials in and around the TPC, which largely occur near its edge.
The position is also used to apply corrections for variations in signal collection efficiencies.
The ratio between S2 and S1 signal sizes is used to discriminate between nuclear recoils (NRs) due to WIMPs or neutron backgrounds, and electronic recoils (ERs) due to $\beta$ or $\gamma$ backgrounds.

Since WIMPs are nonrelativistic~\cite{wimp_review}, the WIMP-nucleus interaction cross section can be written as the sum of a part which increases with the mass of the target nucleus (spin-independent, or SI), and an axial-vector part which couples to the nuclear spin (spin-dependent, or SD) \cite{jungman_review, goodman_witten}.
Recently, XENON1T reported SI results from a ton year exposure, which achieved the lowest ever background in a direct detection experiment and set the most stringent 90\% C.L. upper limit to date on the SI cross section for WIMP masses above 6~\gevcsq~\cite{xenon1t_sr1}.
We also reported the first direct detection constraints on the WIMP-nucleus interaction involving pion exchange currents, an additional scalar component that can dominate if the standard SI interaction is absent or strongly suppressed \cite{wimp_pion}.
In this Letter, we present the first limits on the SD WIMP-nucleon cross sections from the XENON1T experiment, using data from the full tonne year exposure.

\vskip 0.2in

{\it Spin-Dependent Theory.}---The SD interaction of WIMPs with nuclei is described by the WIMP-quark axial-vector--axial-vector Lagrangian~\cite{engel}. At low momentum transfer $q$, the Lagrangian can be evaluated using chiral effective field theory (EFT) \cite{klos_2013, zupan_quark}, and the differential cross section can be written as
\begin{equation}
\frac{d\sigma^{\mathrm{SD}}}{dq^2}=\frac{8G_{F}^{2}}{(2J+1)v^2}S_{A}(q),
\end{equation}
where $G_F$ is the Fermi constant, $v$ is the WIMP velocity in the rest frame of the detector, $J$ is the initial ground-state angular momentum of the nucleus, and $S_A(q)$ is the axial-vector structure factor (all equations shown in natural units). The structure factor is conveniently expressed in terms of the isoscalar ($a_0$) and isovector ($a_1$) WIMP-nucleon couplings as
\begin{equation}
S_A(q) = a_0^2 S_{00}(q) + a_0 a_1 S_{01}(q) + a_1^2 S_{11}(q).\label{eq:2_structure_factor}
\end{equation}
The interaction strength is described by these couplings, while the nuclear structure information is absorbed by the $S_{ij}$ factors. 
The two unknown couplings  ($a_0$, $a_1$) yield a two-dimensional plane of parameter space to search, unlike in the SI case, where the WIMP is typically assumed to have an equal (purely isoscalar) coupling to protons and neutrons.

In the limit of zero momentum transfer, the structure factor simplifies to
\begin{multline}
S_A(0) = \frac{(2J+1)(J+1)}{4\pi J}\\
	\times \lvert (a_0 + a_{1}') \Sp + (a_0 - a_1') \Sn \rvert ^2, \label{eq:3_zero_momentum}
\end{multline}
where $\Sp$ and $\Sn$ are the expectation values of the total proton and neutron spin operators in the nucleus, and $a_1'$ contains a correction to the isovector coupling $a_1$ due to chiral two-body currents involving the exchange of a pion. 
Experimental SD searches constrain the theory using the special cases $a_0=a_1=1$ (``proton only'') and $a_0=-a_1=1$ (``neutron only''). These cases are convenient 
because at the one-body ($a_1' \rightarrow a_1$) level, $S_A(0)$ depends only on the total spin expectation values of the protons and neutrons in the nucleus, respectively~\cite{klos_2013}.

Two naturally occurring isotopes of xenon have nonzero nuclear spin, $^{129}$Xe (spin 1/2) and $^{131}$Xe (spin 3/2), with natural abundances of 26.4\% and 21.2\%, respectively~\cite{crc, natural_abundances}. Residual gas analyzer measurements show consistency with these natural abundances in XENON1T within the 1\% precision of the measurement device. The remaining 52.4\% of xenon has negligible sensitivity to the SD interaction.
As both xenon isotopes have an odd number of neutrons, it follows that $\lvert \Sn \rvert \gg \lvert \Sp \rvert$. 
Specifically, in $^{129}$Xe $\Sn=0.329$ and $\Sp=0.010$, while in $^{131}$Xe $\Sn=-0.272$ and $\Sp=-0.009$~\cite{klos_2013}.
Consequently, XENON1T is more sensitive to the neutron-only case, but also has nonzero sensitivity to the proton-only case since $a_0 - a_1' \neq 0$ in Eq.~\eqref{eq:3_zero_momentum} due to the aforementioned two-body contribution.

The total expected NR spectrum $dR/dE_{\mathrm{r}}$ can be written as
\begin{equation}
\frac{dR}{dE_{\mathrm{r}}}= \frac{2\rho_{\chi}}{m_{\chi}} \int \frac{d\sigma^{\mathrm{SD}}}{dq^2} v f(\vec{v}) d^3 v,\label{eq:4}
\end{equation}
where $m_{\chi}$ is the WIMP mass, $\rho_{\chi}$ is the local WIMP density and $f(\vec{v})$ is the WIMP velocity distribution in the rest frame of the detector. $q = \sqrt{2 E_{\mathrm{r}} m_{\mathrm{Xe}}}$, with $m_{\mathrm{Xe}}$ the mass of a xenon nucleus. A standard isothermal WIMP halo, as in \cite{xenon1t_sr1}, is assumed, with $v_0 = 220~\mathrm{km}/\mathrm{s}$, $\rho_{\chi}~=~0.3~\mathrm{GeV}/(\mathrm{c}^{2}\times\mathrm{cm}^{3})$, $v_\mathrm{esc} = 544~\mathrm{km}/\mathrm{s}$, and Earth velocity $v_\mathrm{E} = 232~\mathrm{km}/\mathrm{s}$ \cite{lewin_smith}. In the neutron- or proton-only case, the differential scattering cross section can be rewritten as
\begin{equation}
\frac{d\sigma^{\mathrm{SD}}}{dq^2}=\frac{\sigmaN}{3 \mu_{N}^{2} v^2} \frac{\pi}{2J+1} S_{N}(q),\label{eq:5_final_diff_cross_section}
\end{equation}
where $\mu_N$ is the reduced mass of the WIMP-nucleon system, $S_N(q)$ is the axial-vector structure factor for a proton or neutron ($N=\{n,p\}$) in xenon (from using the corresponding couplings in Eq.~ \eqref{eq:2_structure_factor}), and $\sigmaN$ is the scattering cross section between a WIMP and a single proton or neutron, at zero momentum transfer~\cite{wimp_pion, jungman_review}. 
Using Eq.~\eqref{eq:5_final_diff_cross_section}, the recoil spectrum in Eq.~\eqref{eq:4} becomes proportional to $\sigmaN$. This unknown parameter is used to set limits as a function of WIMP mass.

Detailed calculations of $S_{N}$ have been carried out in \cite{klos_2013}~for many isotopes relevant to experimental searches, including $^{129}$Xe and $^{131}$Xe. 
These calculations use a detailed nuclear shell model to represent the nuclear states, reproducing the ground-state energies and ordering of energy levels from spectroscopic measurements. Reference~\cite{klos_2013} expands on \cite{menendez_2012}, which was extensively compared with alternative calculations in \cite{xenon100_sd}. 

\begin{figure}[b!]
\centering
\includegraphics[width=\columnwidth]{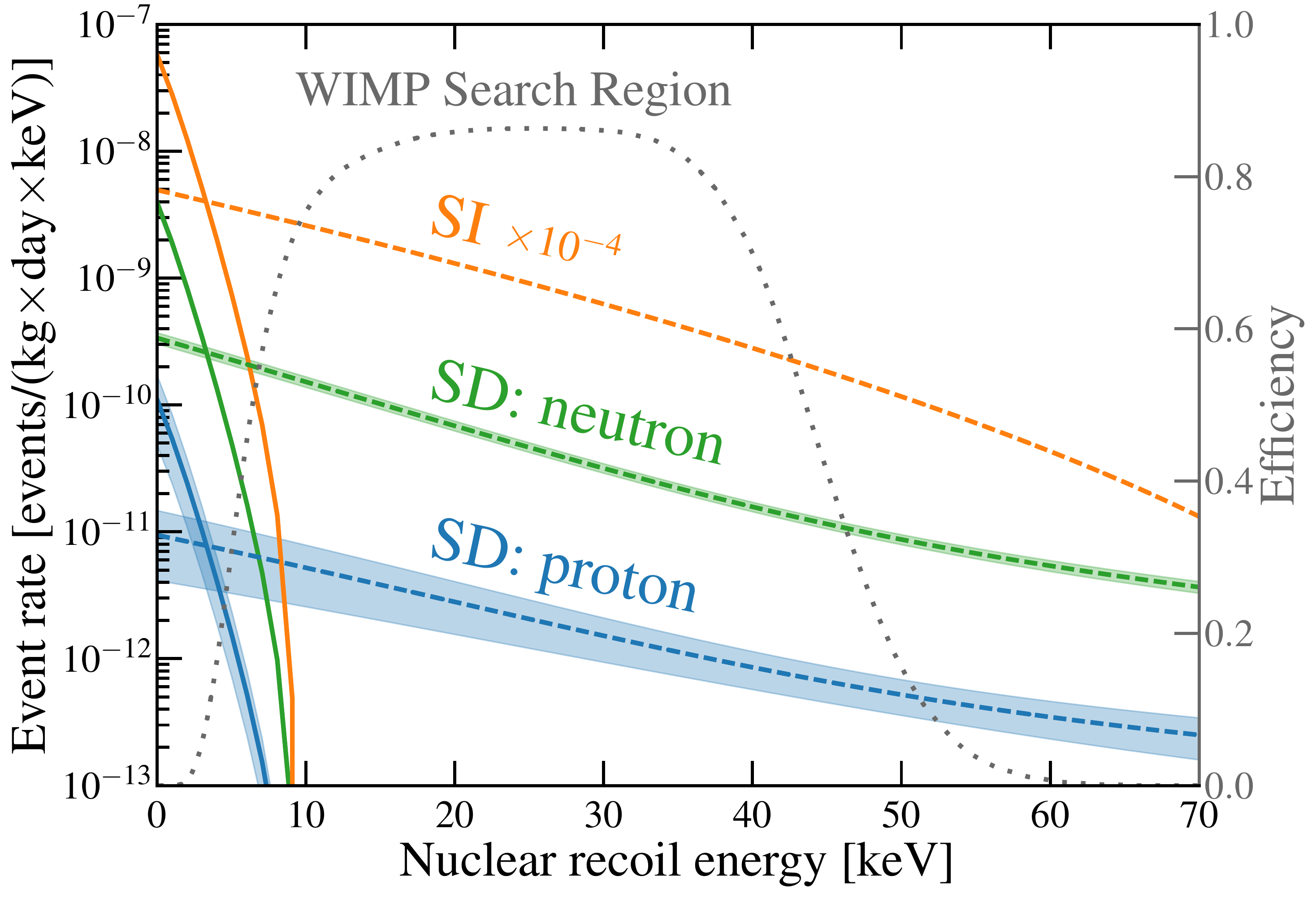}
\caption{\label{fig:1_wimp_spectrum} Comparison of the WIMP-nucleus recoil spectra in the SD neutron-only (green), SD proton-only (blue), and SI (orange, scaled by $10^{-4}$) cases in LXe for a 10~\gevcsq~(solid curve) and 100~\gevcsq~(dashed curve) WIMP and a WIMP-nucleon cross section of $10^{-45}$~cm$^{2}$. 
The bands on the SD spectra come from uncertainties in the contribution to SD scattering from interactions involving the exchange of a pion between two nucleons (two-body currents). The WIMP search region in XENON1T is depicted by the total efficiency curve (gray dotted).}
\end{figure}

The theoretical uncertainties reported in \cite{klos_2013}~mainly come from the two-body current contribution, specifically the density of the nuclear states and the low-energy constants in chiral EFT.
These uncertainties have a larger effect on $\sigmap$ than on $\sigman$, since the SD WIMP-proton sensitivity in a xenon target chiefly relies on two-body interactions with neutrons. 
Since it is difficult to characterize the distribution of these uncertainties, it is conventional \cite{xenon100_sd, lux_sd, pandax_sd} to take the mean of the range of structure factors given, rather than including the uncertainty on the scattering rate in the statistical inference.
Example recoil spectra for the neutron- and proton-only cases are shown in Fig.~\ref{fig:1_wimp_spectrum}, along with the standard SI spectrum (scaled by $10^{-4}$)~\cite{xenon1t_sr1} for reference.
The resulting SD rates are much lower than the SI rates. This is mostly explained by the SD structure factor in Eq.~\eqref{eq:3_zero_momentum}, which is $\mathcal{O}$(1), while the analogous SI form factor scales with the square of the number of nucleons, due to the coherence of the interaction over the nucleus.

\vskip 0.2in

{\it Analysis method.}---SI and SD WIMP-nucleus scattering produce similar recoil spectra in XENON1T, and both interactions produce observables through the same NR process. We therefore use signal corrections and event selection criteria identical to \cite{xenon1t_sr1}.
The dark matter search is limited to events within an inner $1.30~\pm~0.01~\mathrm{t}$ LXe fiducial mass, with corrected S1s between (3, 70)~photoelectrons, accepting NRs of about $5-41~\mathrm{keV}$ nuclear recoil energy on average.
The livetime analyzed is $278.8$~days, consisting of a $32.1$~day run \cite{xe_sr0} and a $246.7$~day run, resulting in a total exposure of $1.0~\mathrm{t} \times \mathrm{yr}$.
Background models, also retained from \cite{xenon1t_sr1}, include data-driven models for accidental coincidence of lone S1s and S2s, and events with reduced charge signal due to interactions at the detector surfaces. 
The ER ($\beta$ and $\gamma$) and NR (radiogenic neutrons and coherent elastic neutrino-nucleus scattering) backgrounds are modeled using energy depositions from GEANT4 simulations, passed to a Monte-Carlo (MC) simulation of ER and NR response in LXe, XENON1T detector physics, and detection efficiency~\cite{analysis_2}. 
The parameters in the MC simulation are determined from a simultaneous fit to calibration data using ER~\cite{xe_rn220} and NR~\cite{rafael_ng} sources taken periodically throughout the exposure. The signal region in the DM search data was blinded prior to the determination of the event selection and background models~\cite{xenon1t_sr1}.
For each WIMP mass, the SD signal recoil spectrum calculated from Eq.~\eqref{eq:4} is propagated through the same MC simulation to generate the expected distribution of S1s and S2s from corresponding WIMP-nucleon interactions.

Statistical inference is done using a three-dimensional (corrected S1, corrected S2 in the bottom PMT array, and radius) unbinned extended likelihood, profiled over nuisance parameters~\cite{analysis_2}.
In addition to these three dimensions, the likelihood distinguishes between events in an inner 0.65 t core and those in an outer section of the fiducial mass to incorporate the difference in the expected neutron background rate, as in \cite{xenon1t_sr1}. 
Nuisance parameters are included to account for uncertainties in ER response, detection and selection efficiencies, and background rates. 
To safeguard against interpreting an under-prediction of ERs as a signal excess, an additional WIMP-like component is added to the background model and constrained by ER calibration data~\cite{safeguard, analysis_2}. 
Upper limits and two-sided intervals are computed using a Feldman-Cousins-based method \cite{feldman_cousins}, with a Neyman band constructed from a profiled likelihood ratio test statistic \cite{pdg_neyman}.
Background-only simulations are performed to calculate the range of possible upper limits under many repetitions of the XENON1T exposure.

\begin{figure}[b]
\centering
\includegraphics[width=\columnwidth]{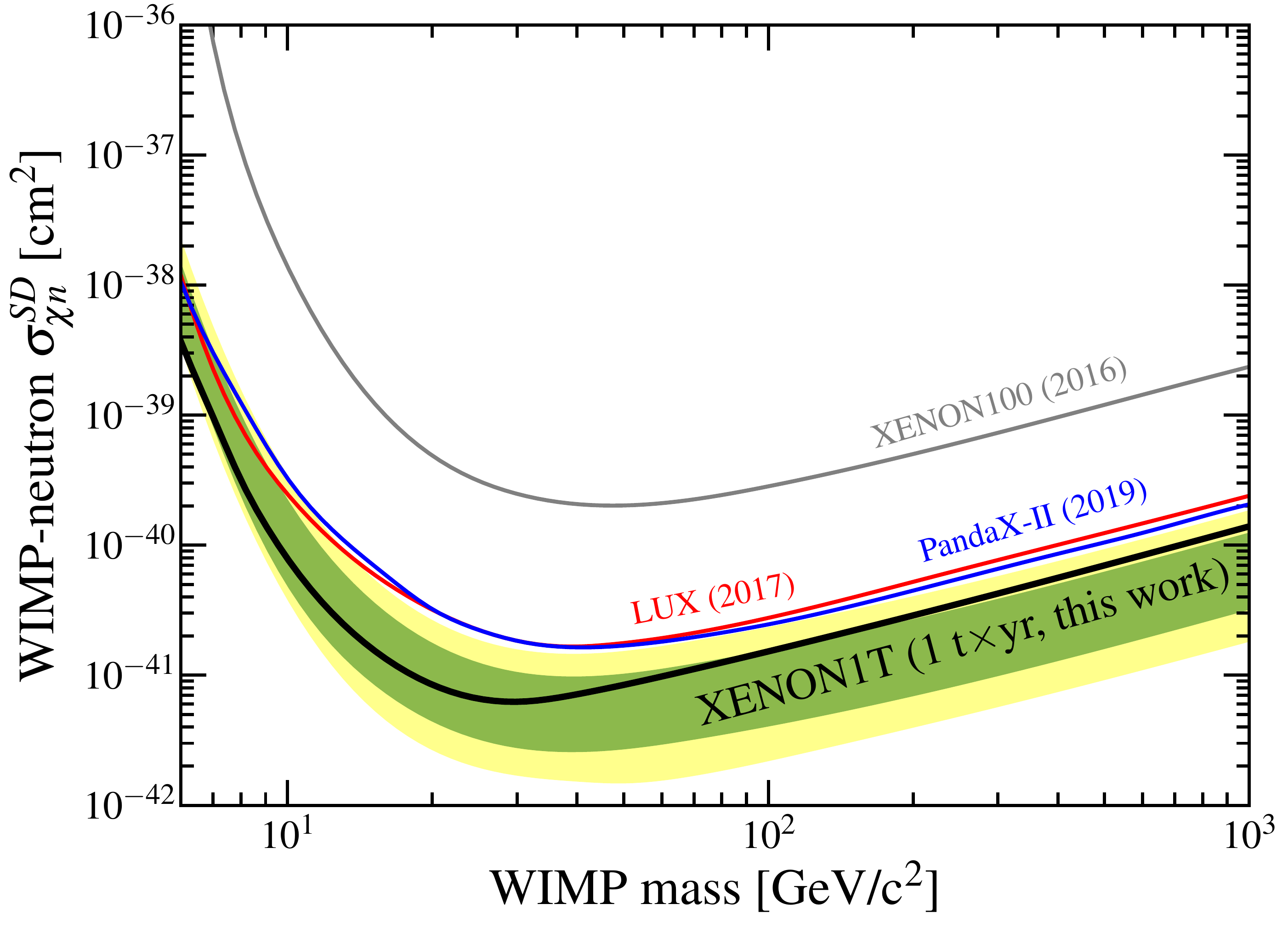}
\caption{\label{fig:2_result} XENON1T 90\% C.L. upper limit on the spin-dependent WIMP-neutron cross section from a 1 ton year exposure. The range of expected sensitivity is indicated by the green (1$\sigma$) and yellow (2$\sigma$) bands. Also shown are the experimental results from XENON100 \cite{xenon100_sd}, LUX \cite{lux_sd} and PandaX-II~\cite{pandax_sd}.
}
\end{figure}
\vskip 0.2in

\begin{figure}[b!]
\centering
\includegraphics[width=\columnwidth]{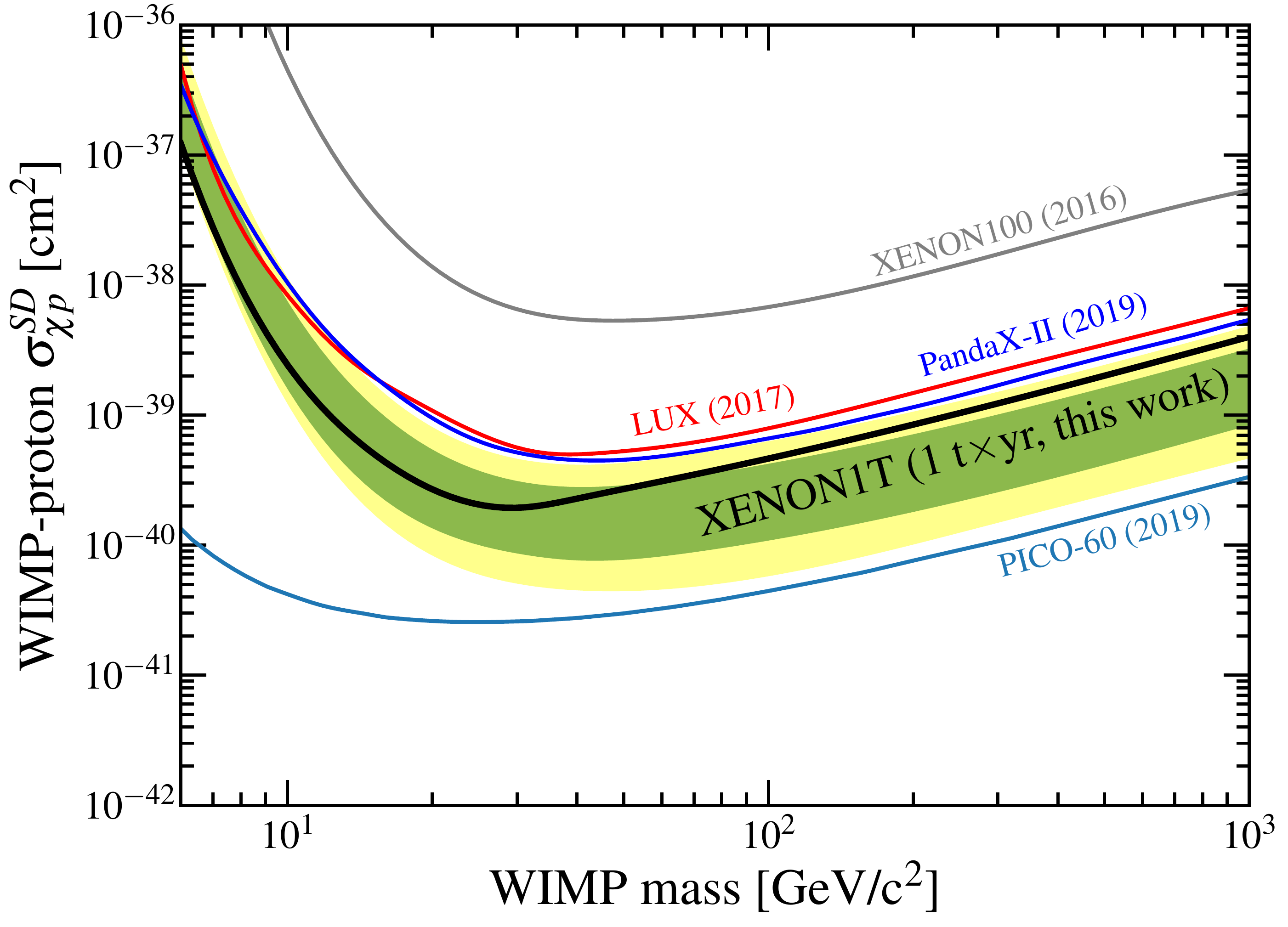}
\caption{\label{fig:3_result} XENON1T 90\% C.L. upper limit on the spin-dependent WIMP-proton cross section from a 1 ton year exposure. The range of expected sensitivity is indicated by the green (1$\sigma$) and yellow (2$\sigma$) bands. Selected experimental results are shown for XENON100 \cite{xenon100_sd}, LUX \cite{lux_sd}, PandaX-II \cite{pandax_sd} and PICO-60 \cite{pico_full}.}
\end{figure}

{\it Results.}---For all WIMP masses considered, and for both the neutron- and proton-only cases, the data are consistent with the background-only hypothesis. 
The local discovery $p$ values at WIMP masses of $6$, $50$, and $200$~\gevcsq~in the neutron-only (proton-only) case are $0.6$, $0.4$, and $0.2$ ($0.6$, $0.3$, and $0.1$), respectively.
Figure~\ref{fig:2_result} and Fig.~\ref{fig:3_result} show the 90\% C.L. upper limits, as well as the 1$\sigma$ and 2$\sigma$ sensitivity bands, on the SD WIMP-neutron and WIMP-proton cross sections, respectively.
Differences between the limit and the median sensitivity due to fluctuation of the background are within the 2$\sigma$ statistical uncertainty.

The mean values of the structure factors are used both for the observed limits and the sensitivity distributions. 
To estimate the impact of the theoretical uncertainty on the result, a cross-check was performed by taking the minimum and maximum values of the structure factors, and using the asymptotic distribution of the test statistic to set limits for each case~\cite{cowan_cranmer}.
At $50$~\gevcsq, the upper limit on the WIMP-neutron cross section shifts downward (upward) by a factor of $1.1$ ($1.1$) when taking the minimum (maximum) structure factor values. Similarly, the upper limit on the WIMP-proton cross section shifts downward (upward) by a factor of $1.6$ ($2.2$) due to the larger dependence of the proton-only sensitivity on the uncertain two-body component.

The neutron-only limit (Fig.~\ref{fig:2_result}) is the most stringent constraint from a direct detection experiment for WIMP masses above 6~\gevcsq, with a minimum of $6.3 \times 10^{-42}~\mathrm{cm}^2$ for a 30~\gevcsq~WIMP.
The proton-only limit (Fig.~\ref{fig:3_result}) is the most stringent constraint from a LXe direct detection experiment, though fluorine-based superheated liquid experiments 
such as PICASSO~\cite{PICASSOFinallimit}, SIMPLE~\cite{SIMPLE}, and PICO-60~\cite{pico,pico_full} have consistently led the field in directly constraining the WIMP-proton cross section. 

Collider experiments are sensitive to WIMPs through searches for final states of {\it pp} collisions with missing transverse energy, which can be attributed to the production of escaping DM particles. 
These searches are complementary to those carried out by direct detection experiments, but direct comparison of the resulting limits requires that a model of DM interactions with standard model particles is specified. 
Following the approach of~\cite{pico}, we use a model recommended by the LHC Dark Matter Working Group~\cite{lhc_comparison}, and frequently used by ATLAS and CMS, to compare results with direct detection SD searches~\cite{atlas, cms}. 
In this model, the WIMP is a Dirac fermion of mass $m_{\chi}$, and has an \mbox{{\it s}-channel} interaction with quarks, mediated by a spin-1 particle of mass $m_{\mathrm{med}}$ with an axial-vector coupling to both the WIMP and the quarks. 
Additionally, the mediator couples equally to all quark flavors, so the WIMP-nucleon interaction is isoscalar. 

Since the two-body pion exchange currents are purely isovector~\cite{klos_2013}, the corresponding correction terms vanish in the isoscalar case. Consequently, the structure factor and, thus, the recoil spectra differ in both shape and rate from the neutron-only and proton-only cases, so this model should be treated as a third, distinct case. 

\begin{figure}[htb!]
\centering  
\includegraphics[width=0.98\columnwidth]{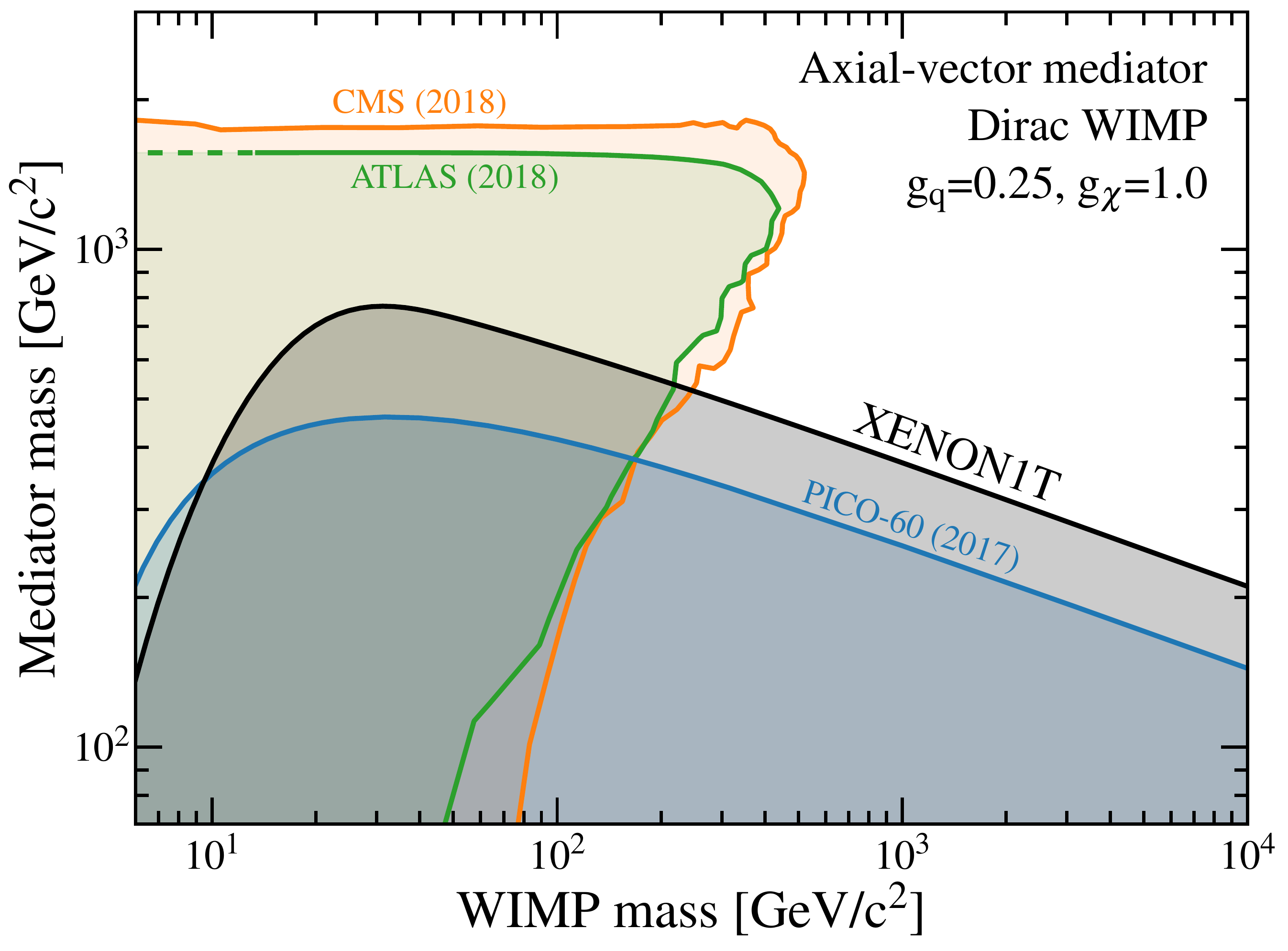}
\caption{\label{fig:4_mediator} XENON1T 95\% C.L. upper limit in the $m_{\mathrm{med}}$-$m_{\chi}$ plane for a simplified isoscalar model with an axial-vector mediator and a Dirac WIMP, where the mediator-quark ($g_q$) and mediator-WIMP ($g_{\chi}$) couplings are fixed to 0.25 and 1.0, respectively. Shown for comparison are 95\% C.L. limits from PICO-60~\cite{pico}, ATLAS~\cite{atlas}, and CMS~\cite{cms}. The shaded regions are excluded by the correspondingly colored limits.}
\end{figure}

The model contains four free parameters: the mediator mass $m_{\mathrm{med}}$, $m_{\chi}$, the mediator-quark coupling $g_q$, and the mediator-WIMP coupling $g_{\chi}$. 
Following~\cite{lhc_comparison}, the cross section can now be written as
\begin{equation}
    \sigman = \sigmap = \frac{0.31}{\pi} \frac{g_q^2 g_{\chi}^2 \mu_N^2}{m_{\mathrm{med}}^4}. \label{eq:6_mediator}
\end{equation}
New signal models are generated, and statistical inference is performed via the same method as for the neutron-only and proton-only cases, resulting in a 95\% C.L. upper limit on the WIMP-nucleon cross section $\sigmaN$ for the isoscalar case.
After substituting the conventional values of the couplings ($g_q=0.25$, $g_{\chi}=1$), we can use Eq.~\ref{eq:6_mediator} to transform this upper limit into the $m_{\mathrm{med}}$-$m_{\chi}$ plane, and compare it directly with collider experiments for this particular simplified model. 
As shown in Fig.~\ref{fig:4_mediator}, the constraint from XENON1T data excludes new parameter space in this theory, and represents the most stringent constraint from a direct detection experiment. 
We note that in a complete treatment of the comparison between the LHC and direct detection experiments, one should include isospin-violating corrections to the couplings due to the difference in energy scales~\cite{panci}. This cannot be simply included because the structure factor becomes dependent on the mediator mass. As this effect would enhance the rate of WIMP-neutron scattering relative to WIMP-proton scattering, our limit can be viewed as conservative.

\vskip 0.2in
{\it Conclusion.}---
We have analyzed the data from a ton year exposure of XENON1T, which is currently the leading dual-phase liquid xenon direct detection experiment. The SD WIMP-neutron and WIMP-proton spectra were calculated from the work of Ref.~\cite{klos_2013}. For the data selection, background models and statistical interference, the same methods have been used as for the XENON1T SI analysis \cite{xenon1t_sr1}. The data are consistent with the background-only hypothesis and the resulting limits were computed. These results are the first constraints on the SD interaction from XENON1T and improve upon previous constraints from LXe experiments.
Additionally, the XENON1T 95\% C.L. upper limit in the $m_{\mathrm{med}}$--$m_{\chi}$ plane for a simplified isoscalar model has been compared with constraints from collider searches and excludes new parameter space. 
This Letter is part of a program to constrain a large set of theoretical parameters using XENON1T, including the SI \cite{xenon1t_sr1} and WIMP-pion \cite{wimp_pion} interactions, along with future work to constrain a broader set of WIMP-quark couplings using EFT methods.

\vskip 0.2in

{\it Acknowledgements.}---
The authors would like to thank A. Schwenk, M. Hoferichter, and J. Men\'endez for useful discussions on SD theory and nuclear structure.
We gratefully acknowledge support from the National Science Foundation, Swiss National Science Foundation, German Ministry for Education and Research, Max Planck Gesellschaft, Deutsche Forschungsgemeinschaft, Netherlands Organisation for Scientific Research (NWO), Netherlands eScience Center (NLeSC) with the support of the SURF Cooperative, Weizmann Institute of Science, Israeli Centers Of Research Excellence (I-CORE), Pazy-Vatat, Initial Training Network Invisibles (Marie Curie Actions, PITNGA-2011-289442), Fundacao para a Ciencia e a Tecnologia, Region des Pays de la Loire, Knut and Alice Wallenberg Foundation, Kavli Foundation, and Istituto Nazionale di Fisica Nucleare. Data processing is performed using infrastructures from the Open Science Grid and European Grid Initiative. We are grateful to Laboratori Nazionali del Gran Sasso for hosting and supporting the XENON project.

\end{document}